\documentclass[10pt,5p,twocolumn]{elsarticle}

\usepackage{graphicx}
\usepackage{amsmath,amssymb,bm}
\usepackage{multirow}
\biboptions{sort&compress,comma}
\journal{Surface Science}

\begin{document}
\begin{frontmatter}

\title{Phonons of graphene and graphitic materials derived from the empirical potential LCBOPII}
\author[ru]{L.J.~Karssemeijer}
\ead{lkarssem@science.ru.nl}
\author[ru]{Annalisa~Fasolino\corref{cor1}}
\ead{a.fasolino@science.ru.nl}
\address[ru]{Institute for Molecules and Materials, Radboud University Nijmegen, Heyendaalseweg 135, 6525 AJ Nijmegen, The Netherlands}
\cortext[cor1]{Corresponding author}

\begin{abstract}
We present the interatomic force constants and phonon dispersions of graphite and graphene from the LCBOPII empirical bond order potential. We find a good agreement with experimental results, particularly in comparison to other bond order potentials. From the flexural mode we determine the bending rigidity of graphene to be 0.69 eV at zero temperature. We discuss the large increase of this constant with temperature and argue that derivation of force constants from experimental values should take this feature into account.  We examine also other graphitic systems, including multilayer graphene for which we show that the splitting of the flexural mode can provide a tool for characterization.
\end{abstract}

\begin{keyword}
	phonons \sep graphene \sep graphite \sep force constants \sep LCBOPII \sep bending rigidity
\end{keyword}

\end{frontmatter}

\section{Introduction}
The phonon spectrum of a crystalline solid provides information on several important  physical properties like sound velocities, thermal conductivity, heat capacity and thermal expansion. The phonon spectrum of graphite has been intensively studied experimentally~\cite{Maultzsch2004,Siebentritt1997,Mohr2007,Nicklow1972,Oshima1988,Tuinstra1970} and theoretically~\cite{Dresselhaus,Rouffignac1981,Al-Jishi1982,Benedek1993} in the past and some models have also been shown to be accurate for the description of fullerenes~\cite{Benedek1993} and of graphite slabs~\cite{Rouffignac1981}. In the more recent past, Raman spectroscopy  has  proven to be of crucial importance also for the characterization of graphene and nanotubes as well as for graphitic nanostructures of lower symmetry, like bent tubes and graphene edges~\cite{Ferrari2006,Casiraghi2009,Calizo2007}. However, the  many unusual structural aspects of graphene, like the observed ripples~\cite{Meyer2007}, negative thermal expansion~\cite{Bao2009,Zakharchenko2009}, edge reconstruction~\cite{Koskinen2009} and localized~\cite{Carlsson2006} and extended defects~\cite{Lahiri2010,Coraux2008} make it desirable to describe the energetics of carbon in different structural and bonding  configurations beyond the harmonic approximation by means of a unique potential.  Bond order potentials are a class of empirical interatomic potentials (EIPs) designed for this purpose~\cite{Tersoff1986,Brenner1990,Los2003}. They aim at describing not  only the structure around equilibrium but also anharmonic effects~\cite{Fasolino2007} and the possible breaking and formation of bonds in structural phase transitions like the graphite to diamond transition where the character of the bonding changes from  sp$^2$ to sp$^3$~\cite{Glosli1999,Ghiringhelli2005}. In view of this larger and exacting scope it may be expected that the phonon spectra derived from these potentials are not as accurate as those derived from models meant to describe a single specific situation. However they allow to study, without further adjustment of parameters,  all carbon structures, including the effect of defects, edges and other structural changes, also as a function of temperature. The purpose of this paper is to evaluate the force constants and phonon spectrum of graphitic structures derived from the  Long-range Carbon Bond Order Potential (LCBOPII)~\cite{Los2003,Los2005} and compare these results to experimental values, force constant models and to the Tersoff~\cite{Tersoff1988} and Brenner~\cite{Brenner1990} bond order potentials for carbon.

The phonon dispersions of graphene and graphite have been measured experimentally~\cite{Maultzsch2004,Siebentritt1997,Mohr2007,Nicklow1972,Oshima1988,Yanagisawa2005,Tuinstra1970}, determined from \textit{ab initio} calculations~\cite{Mounet2005,Maultzsch2004,Wirtz2004,Dubay2003} and calculated from bond order EIPs~\cite{Lindsay2010,Tewary2009}. The \textit{ab initio} results generally agree very well with experimental measurements whereas widely used EIPs such as the Tersoff~\cite{Tersoff1988} and Brenner~\cite{Brenner1990} potentials give less accurate results~\cite{Lindsay2010}, particularly in the optical region. One reason for this is that the range of interatomic interactions in EIPs is limited for computational efficiency whereas force constant models show that interatomic force constants (IFCs) up to fourth or even fifth nearest neighbours (NNs) must be included for accurate phonon dispersions~\cite{Wirtz2004,Mohr2007}. The second generation LCBOPII~\cite{Los2005} EIP includes long-range interactions up to 6 \AA\ which is well beyond fifth NNs in graphene and it is interesting to study their effect on the phonons in comparison to other approaches. 

\begin{figure*}
	\begin{center}
	\includegraphics{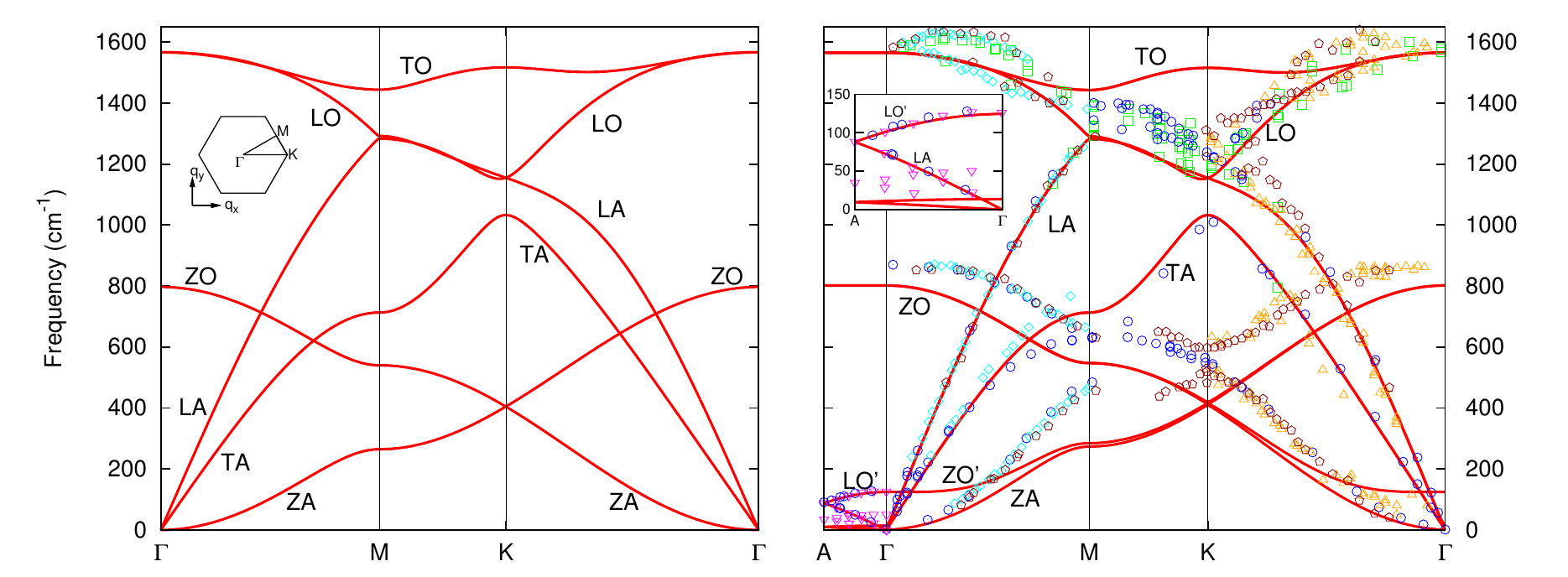}
	\caption{Phonon frequency in cm$^{-1}$. Left: graphene phonon dispersion from LCBOPII; Right: Graphite phonon dispersion from LCBOPII with experimental data, the inset is an enlargement of the low-frequency dispersion along the $\bm{A}-\bm{\Gamma}$ line. The locations of the high symmetry points are $\bm{M}=\pi/\sqrt{3}a(\sqrt{3},1,0)$, $\bm{K}=4\pi/3a(1,0,0)$ and $\bm{A}=\pi/c(0,0,1)$ in the coordinate system defined in the inset of the left figure. The experimental data for graphite are from Ref.~\cite{Maultzsch2004} (squares), Ref.~\cite{Mohr2007} (circles), Ref.~\cite{Siebentritt1997} (triangles), Ref.~\cite{Oshima1988} (diamonds), Ref.~\cite{Nicklow1972} (inverse triangles) and Ref.~\cite{Yanagisawa2005} (pentagons).}
	\label{fig1}
	\end{center}
\end{figure*}
In Section~\ref{sec:methods} we describe the computational method with emphasis on the anomalous dispersion of the flexural mode. In Section~\ref{sec:phonon} we present the LCBOPII phonon dispersion of graphene and graphite, compare our force constants to other models and examine the role of specific force constants on the phonon dispersions. We devote Section~\ref{sec:bendingrigidity} to the analysis of the bending rigidity and its temperature dependence. In Section~\ref{sec:other} we show the phonons of (10,10) nanotubes and show the relevance of low-energy phonons of multilayer graphene for their characterization.

\section{Methods}
\label{sec:methods}
The phonon dispersions are calculated by means of standard lattice dynamics~\cite{Born}.
The interatomic force constants are calculated by evaluating, by central differences, the second derivatives (the IFCs) of the LCBOPII EIP with respect to atomic displacements around their equilibrium positions. 
The phonon frequencies at wavevector $\bm{q}$, $\omega(\bm{q})$, are determined by diagonalizing the dynamical matrix 
\begin{equation}
 	D_{\alpha,\beta}^{k,k'}(\bm{q})= \frac{1}{\sqrt{m_k m_{k'}}}\sum_{\bm{R}}\phi_{\alpha,\beta}^{k,k'}(\bm{R}) e^{i\bm{q}\cdot\bm{R}},
\end{equation}
 where $\phi_{\alpha,\beta}^{k,k'}(\bm{R}) $ is the force constant matrix, $\alpha,\beta$ being Cartesian indices, for two atoms $k$ and $k'$ in unit cells separated by a lattice vector $\bm{R}$. 
In layered materials the lowest, out-of-plane, acoustic phonon branch (ZA) has a peculiar quadratic dispersion near the zone center with a coefficient determined by the bending rigidity $\kappa$ of the crystal. For graphite, it has been shown by Lifshitz~\cite{Lifshitz1952} that the dispersion has the following form:
\begin{equation}
	\omega_{\textrm{ZA}}(\bm{q})=\sqrt{\frac{C_{44}}{\rho_{3D}}|\bm{q}|^2+\frac{\kappa}{\rho_{3D} c}|\bm{q}|^4},
\label{zadispersiongraphite}
\end{equation}
where $\rho_{3D}=8m_C/(\sqrt{3}a^2 c)$ is the mass density, $C_{44}$ is the shear elastic constant, $c$ is the lattice parameter equal to twice the interlayer distance in $ABAB$ stacked graphite, $a$ is the in-plane lattice parameter and $m_C$ is the atomic mass of carbon. For graphene, the dispersion reduces to a purely quadratic form:
\begin{equation}
	\omega_{\textrm{ZA}}(\bm{q})=\sqrt{\frac{\kappa}{\rho_{2D}}}|\bm{q}|^2,
\label{zadispersiongraphene}
\end{equation}
where $\rho_{2D}=4m_C/(\sqrt{3}a^2)$ is now a two-dimensional mass density.

\begin{table*}
\begin{center}
\caption{Graphene phonon frequencies from LCBOPII at high symmetry points in cm$^{-1}$. Experimental values for graphite are also listed: $^a$~Ref.~\cite{Nicklow1972}, $^b$~Ref.~\cite{Oshima1988}, $^c$~Ref.~\cite{Maultzsch2004}, $^d$~Ref.~\cite{Yanagisawa2005} (graphene), $^e$~Ref.~\cite{Siebentritt1997}, $^f$~Ref.~\cite{Tuinstra1970} and $^g$~Ref.~\cite{Mohr2007}. 
}
\label{table1}
\vspace{0.2cm}
\begin{tabular}{|l|cc|cc|cc|}
\hline
Mode & \multicolumn{2}{|c|}{$\bm{\Gamma}$} & \multicolumn{2}{|c}{$\bm{M}$} & \multicolumn{2}{|c|}{$\bm{K}$} \\
	& LCBOPII & Experiment &  LCBOPII & Experiment &  LCBOPII & Experiment \\
\hline\hline
ZA 	& 0 	& 						& 265 & 471$^a$, 465$^b$, 451$^d$, 485$^g$		& 405 &	482$^d$, 517$^d$, 530$^e$, 540$^g$\\
TA 	& 0 	& 						& 713 & 630$^d$, 631$^g$					& 1033 & 	1010$^g$		\\
LA 	& 0 	& 						& 1282 & 1290$^c$						& 1153 &	1194$^c$, 1224$^h$	\\
ZO 	& 797  & 861$^b$, 868$^g$		& 540 & 670$^b$, 631$^g$					& 405 &	588$^d$, 627$^e$, 540$^g$\\
LO 	& 1563 & 1590$^b$, 1575$^f$, 1582$^d$	& 1290 & 	1323$^c$			& 1153 &	1194$^c$, 1224$^d$\\
TO 	& 1563 & 1590$^b$, 1575$^f$, 1582$^d$	& 1441 &	1390$^c$, 1389$^b$		& 1513 &	1310$^d$, 1291$^e$\\
\hline
\end{tabular}
\end{center}
\end{table*}

\section{Phonon dispersion}
\label{sec:phonon}

Minimization of the LCBOPII cohesive energy with respect to the lattice parameters gives $a=\sqrt{3}a_{CC}=2.4592$~\AA\ for graphene and $a=\sqrt{3}a_{CC}=2.4585$~\AA, $c=6.7344$~\AA\ for graphite. The phonon dispersions for graphene and $ABAB$ graphite, calculated at these lattice parameters, are shown in Fig.~\ref{fig1}.
The branches are classified as follows: L stands for longitudinal in-plane, T for transverse in-plane and Z for transverse out-of-plane polarization at the $\bm{\Gamma}$ point. An A refers to acoustic modes and an O to optical
modes. The O' modes in graphite indicate out-of-phase oscillation of two equivalent atoms in neighbouring layers.
The phonon dispersions of graphite and graphene are very similar due to the weakness of the interlayer interactions compared to the strong covalent bonds binding the atoms in the layers. Consequently, most of the twelve branches in graphite are almost doubly degenerate with the exception of the out-of-plane branches below 400 cm$^{-1}$.

Contrary to the two  linear, in-plane acoustic LA and TA, modes the out-of-plane ZA mode has a quadratic dispersion near $\bm{\Gamma}$ which is typical of layered crystals~\cite{Zabel2001}. The ZA mode is a bending mode, the two atoms in the unit cell move in phase in the $z$-direction, which, at long wavelengths, bends the surface resulting in rippling of the graphene sheet. The softness of this mode also means that it plays a dominant role at low temperatures.
Also the optical out-of-plane, ZO mode has a considerably lower energy than the other optical branches due to the fact that atoms are much more free to move perpendicular to the plane than in the plane itself.
At the $\bm{K}$-point, the TO/LO and the ZA/ZO modes are degenerate by symmetry. 

In Fig.~\ref{fig1} and Table~\ref{table1} we compare the LCBOPII phonon spectrum to experimental results by high resolution electron energy-loss spectroscopy (HREELS)~\cite{Siebentritt1997,Oshima1988,Yanagisawa2005}, inelastic x-ray scattering~\cite{Maultzsch2004,Mohr2007} and inelastic neutron scattering~\cite{Nicklow1972}. The overall agreement with the experimental values is rather good, considering that the potential was not specifically fitted to reproduce the force constants of graphite.

LCBOPII performs very well compared to the popular Tersoff~\cite{Tersoff1988} and Brenner~\cite{Brenner1990} EIPs, for which the phonon dispersions were recently published~\cite{Lindsay2010}. 
The Tersoff EIP overestimates the LO and TO branches by nearly 40\% and both potentials show large discrepancies with experiments in the in-plane acoustic branches which are very well reproduced by LCBOPII. The latter modes are of particular importance for the thermal conductivity in graphene~\cite{Nika2009}.
The only deviation occurs at the $\bm{M}$ point for the TA branch, where LCBOPII overestimates the experimental value from Ref.~\cite{Mohr2007} by 13\%. The measurements from Ref.~\cite{Oshima1988} show even higher frequencies for this mode but these may have been obtained from a sample of poor quality as HREELS selection rules state that the TA mode should not be observable along the $\bm{\Gamma}-\bm{M}$ line~\cite{Yanagisawa2005,Mohr2007}. \textit{Ab initio} calculations also confirm the experimental results from Ref.~\cite{Mohr2007}. From the slopes of the TA and LA modes we determine their respective sound velocities as 13.0 and 20.7 km/s which compare well to the experimental values of 14.7 and 22.2 km/s~\cite{Bosak2007}.

The quadratic dispersion of the ZA mode is reproduced well by LCBOPII, but the frequency is underestimated. We argue that this mode is strongly temperature dependent as we discuss in Section~\ref{sec:bendingrigidity}. Also the ZO mode is found to be lower than all experiments, the difference being about 8\% at $\bm{\Gamma}$.

The low-energy dispersion of graphite for wavevectors parallel to the $c$-axis is shown in the inset of Fig.~\ref{fig1}. In these modes, layers oscillate rigidly and frequencies are thus determined by the long-range interactions of LCBOPII. The two longitudinal, LA and LO', `breathing' modes are in excellent agreement with experiments. This comes as no surprise since the compressibility of graphite was one of the parameters used in fitting the long-range interactions~\cite{Los2005}. The lower, doubly degenerate, transverse branches, corresponding to the shearing motion of the layers are instead too soft compared to experiments as we discuss later in Section~\ref{sec:bendingrigidity}.

The description of the highest optical branches requires long-range IFCs up to fourth or fifth NNs~\cite{Wirtz2004,Mohr2007} and are therefore the most difficult to reproduce with EIPs~\cite{Lindsay2010}. LCBOPII includes interactions in this range through its long-range potential, but these are merely pair interactions of Morse form which are too smooth to produce significant force constants.
In particular, the flat dispersion of the TO mode along the $\bm{M}-\bm{K}$ line differs from experiments and \textit{ab initio} calculations. The difference reaches 15\% at the $\bm{K}$-point. Another missing feature is the overbending of the LO mode, namely the shift of the highest frequencies away from $\bm{\Gamma}$. This overbending is believed to originate from strong electron-phonon interactions which lower the frequencies of the highest phonon modes at $\bm{\Gamma}$ and $\bm{K}$~\cite{Piscanec2004,Maultzsch2004}. For these branches the Brenner EIP gives a behaviour similar to LCBOPII.

\begin{table*}
\begin{center}
\caption{Force constants for graphene from LCBOPII compared to force constant models from Refs.~\cite{Mohr2007,Wirtz2004,Tewary2009} which were fitted to reproduce experimental results.}
\label{table2}
\begin{tabular}{|c|r@{.}l r@{.}l r@{.}l r@{.}l|r@{.}l r@{.}l r@{.}l r@{.}l|r@{.}l r@{.}l r@{.}l r@{.}l|}
\hline
$i$ & \multicolumn{8}{|c|}{Stretching, $\phi^{(i)}_{\textrm{st}}$ (eV/\AA$^2$)} & \multicolumn{8}{|c|}{In-plane, $\phi^{(i)}_{\textrm{ip}}$ (eV/\AA$^2$)} & \multicolumn{8}{|c|}{Out-of-plane, $\phi^{(i)}_{\textrm{z}}$ (eV/\AA$^2$)}\\
(Distance)	& \multicolumn{2}{c}{LCBOPII}& \multicolumn{2}{c}{\cite{Mohr2007}} & \multicolumn{2}{c}{\cite{Wirtz2004}} & \multicolumn{2}{c|}{\cite{Tewary2009}} & \multicolumn{2}{c}{LCBOPII} & \multicolumn{2}{c}{\cite{Mohr2007}} & \multicolumn{2}{c}{\cite{Wirtz2004}} & \multicolumn{2}{c|}{\cite{Tewary2009}} & \multicolumn{2}{c}{LCBOPII} & \multicolumn{2}{c}{\cite{Mohr2007}} & \multicolumn{2}{c}{\cite{Wirtz2004}} & \multicolumn{2}{c|}{\cite{Tewary2009}} \\
\hline\hline
1 (1.42 \AA) & 26&60 & 25&88 & 25&58 & 25&57 & 8&99 & 8&42 & 9&05 & 9&05 & 4&73 & 6&18 & 6&17 & 6&17 \\
2 (2.46 \AA) & 3&37 & 4&04 & 4&63 & -2&55 & -0&61 & -3&04 & -2&55 & 4&63 & -0&75 & -0&49 & -0&51 & -0&51 \\
3 (2.84 \AA) & 0&51 & -3&02 & -2&07 & -2&07 &-0&05 & 3&95 & 3&13 & 3&13 & -0&05 & 0&52 & 0&36 & 0&36 \\
4 (3.76 \AA) & 0&02 & 0&56 & 0&41 & 0&66 & 0&00 & 0&13 & 0&34 & 0&31 & 0&00 & -0&52 & -0&32 & -0&33 \\
5 (4.26 \AA) & 0&00 & 1&03 & 0&00 & 0&00 & 0&00 & 0&11 & 0&00 & 0&00 & 0&00 & 0&17 & 0&00 &0&00 \\
\hline
\end{tabular}
\end{center}
\end{table*}
To gain more insight in the origin of the discrepancies with experiments we compare in Table~\ref{table2} the force constants of LCBOPII to IFC sets proposed in Refs.~\cite{Mohr2007,Wirtz2004,Tewary2009}. The coordinate system is chosen such that $x$ is the coordinate along the line connecting two atoms, $y$ is the in-plane coordinate perpendicular to this direction and $z$ is the coordinate perpendicular to the plane. The IFCs between $i$-th NNs in these directions are respectively the bond stretching, $\phi^{(i)}_{\textrm{st}}$, the transverse, $\phi_{\textrm{tr}}^{(i)}$, and the out-of-plane, $\phi^{(i)}_{\textrm{z}}$, force constants.  The general form of the force constant matrix for the $i$-th NNs in graphene is
\begin{equation}
\begin{pmatrix}
\phi_{\textrm{st}}^{(i)}	& \phi_{\textrm{od}}^{(i)}	& 0 \\
-\phi_{\textrm{od}}^{(i)}	& \phi_{\textrm{tr}}^{(i)} & 0 \\
0		&	0	& \phi_{\textrm{z}}^{(i)} \\
\end{pmatrix},
\label{eq:fcmatrix}
\end{equation}
where the off-diagonal, $\phi^{(i)}_{\textrm{od}}$, elements for $i=1,3,5$ are equal to zero due to the hexagonal symmetry, $\phi^{(2)}_{\textrm{od}}=1.48$ eV/\AA$^2$ and $\phi^{(5)}_{\textrm{od}}=\mathcal{O}(10^{-6})$ eV/\AA$^2$ for LCBOPII.
The IFCs from Ref.~\cite{Mohr2007} are obtained by fitting a fourth NN force constants model to experimental values, those from Ref.~\cite{Tewary2009} are derived from an extended Brenner EIP, also fitted to experimental values, and the IFCs from Ref.~\cite{Wirtz2004} are from a fourth NN force constant model including a nonzero $\phi_{\textrm{od}}^{(2)}$ of -0.57 eV/\AA$^2$ fitted to the dispersion obtained \textit{ab initio} within the DFT-GGA approximation. 

From the comparison of Table \ref{table2} we see that the first NN IFCs are very similar to the fitted force constants of the reference models. However the decay of the LCBOPII force constants beyond first NNs is too rapid compared to the sets of IFCs which reproduce the experimental values accurately.

\begin{figure}[h]
\begin{center}
\includegraphics{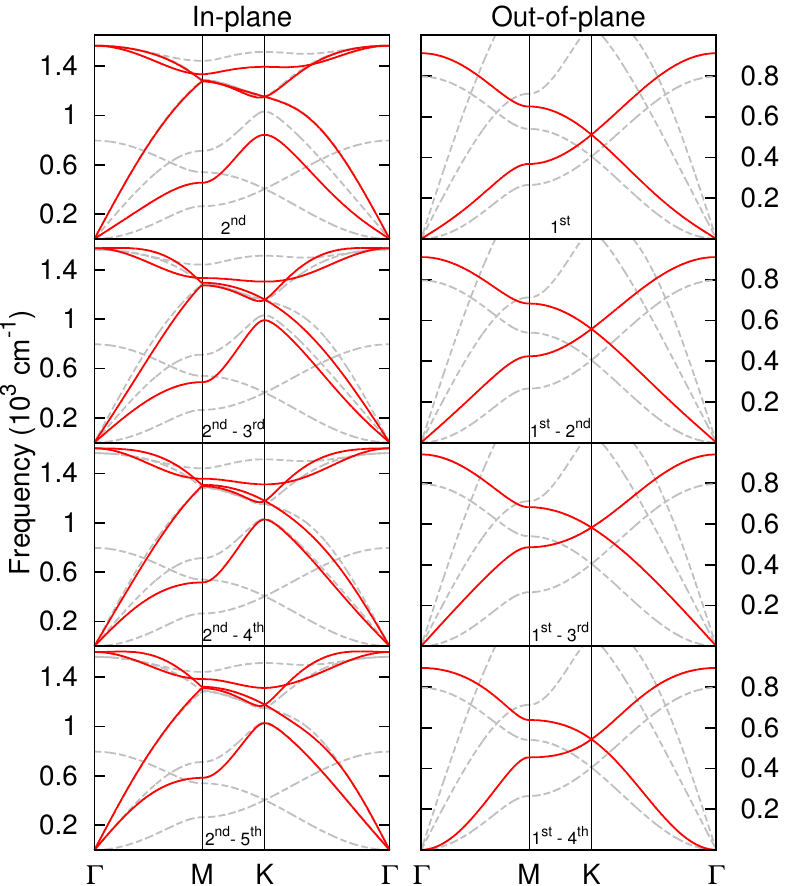}
\caption{Left panels: modification of the in-plane force constants from the second to the indicated level of NNs to match those from Ref.~\cite{Mohr2007}; Right panels: modification of the out-of-plane force constants from the first to the indicated level of NNs to match those from Ref.~\cite{Tewary2009}. Notice the different scales. Red (solid) lines are the branches that are modified as a consequence of the change in force constants. The gray (dashed) lines are the original LCBOPII dispersions.}
\label{fig2}
\end{center}
\end{figure}

To see how the in-plane phonon branches evolve if larger IFCs beyond first NNs are included we manually increase these force constants. We consecutively changed the in-plane IFCs from second to fifth NNs to match those from Ref. \cite{Mohr2007}. Since $\phi^1_{z}$ is already 23\% lower than the fitted values we changed the out-of-plane IFCs from first NNs. They were matched to those obtained in Ref. \cite{Tewary2009} since their model includes interactions up to fourth NNs only which better resembles LCBOPII. 

The resulting phonon dispersions are shown in Fig.~\ref{fig2}. The modification of the in-plane IFCs beyond first NNs clearly improves the phonon dispersion. Changes up to third NNs lower the frequencies of the transverse branches, particularly the TO branch along $\bm{M}-\bm{K}$ and the TA branch at the $\bm{M}$-point but perturbs the good agreement of the sound velocities of the linear modes. With in-plane IFCs changed up to fifth NNs the dispersion is in excellent agreement with experiments. This means that an improvement of only the long-range interactions of LCBOPII can considerably improve the accuracy of the potential for graphitic systems. However, this is not an easy task, since in the construction of LCBOPII short and long range interactions are fitted simultaneously. For the out-of-plane branches the optical ZO branch is greatly improved by the increase of the first NN IFC but the important quadratic behaviour of the ZA mode is lost. Interestingly the quadratic dispersion is recovered only once fourth NNs interaction is included.

\section{Quadratic ZA dispersion and bending rigidity}
\label{sec:bendingrigidity}
The bending rigidity $\kappa$ is the key quantity which characterizes the mechanical properties of membranes~\cite{Nelson}.
For a crystalline membrane like graphene it is intimately related to the quadratic ZA branch through Eq.~\eqref{zadispersiongraphene}. Reported values of $\kappa$ vary from 0.79 to 2.13 eV~\cite{Tersoff1992,Tewary2009,Fasolino2007,Zakharchenko2010,Perebeinos2009,Zhanchun2002}. Besides the different techniques used to calculate $\kappa$ and the different models of carbon interactions there might be other reasons for the confusing variety of reported values. First, when comparing the values of $\kappa$ for graphite and graphene, one should consider the bending rigidity \emph{per layer} and not per unit cell since the latter results in a factor two difference. In fact, since graphite has two graphene layers in the unit cell, the coefficients of the $|\bm{q}|^2$ term in Eqs.~\eqref{zadispersiongraphene} and~\eqref{zadispersiongraphite} differ by a factor two while (see Fig.~\ref{fig1}) the quadratic coefficients should be approximately equal.
The second, more important reason, is that the bending rigidity of graphene has been found to be strongly temperature dependent in detailed Monte Carlo simulations~\cite{Fasolino2007,Zakharchenko2010}. Contrary to liquid membranes~\cite{Nelson}, $\kappa$ increases with increasing temperature. This increase reaches roughly 40\% already at room temperature~\cite{Fasolino2007,Zakharchenko2010}.  The temperature dependence of the bending rigidity implies that also the ZA phonon mode should depend on temperature which makes comparison of zero temperature dispersion as presented here to the room temperature experimental values non straightforward. 

From the fit of Eq.~\eqref{zadispersiongraphene} to the ZA dispersion along the $\bm{\Gamma}-\bm{M}$ line we obtain for the bending rigidity $\kappa=(0.69\pm 0.02)$ eV for graphene. The same procedure for graphite with Eq.~\eqref{zadispersiongraphite} yields $\kappa=(0.69\pm0.01)$ eV per layer and $C_{44}=(5.8\pm0.02) \times 10^8$ Pa. This value of $C_{44}$ is much lower than the experimental value of $5.03\times10^9$ Pa~\cite{Bosak2007} due to the too small corrugation energy of LCBOPII which gives a difference of about 1.5 meV/atom against about 10 meV/atom~\cite{Kolmogorov2005,Savini2010}between $AA$ and $AB$ graphite stacking. As a consequence, the transverse modes of graphite for wavevectors parallel to the $c$-axis shown in the inset of Fig.~\ref{fig1}, have too low frequencies. The bending rigidities of graphite and graphene per layer are the same, which is in agreement with the results from Ref.~\cite{Zakharchenko2010} who find almost equal bending rigidities per layer for graphene and bilayer graphene. 

\begin{figure}[h]
\begin{center}
\includegraphics{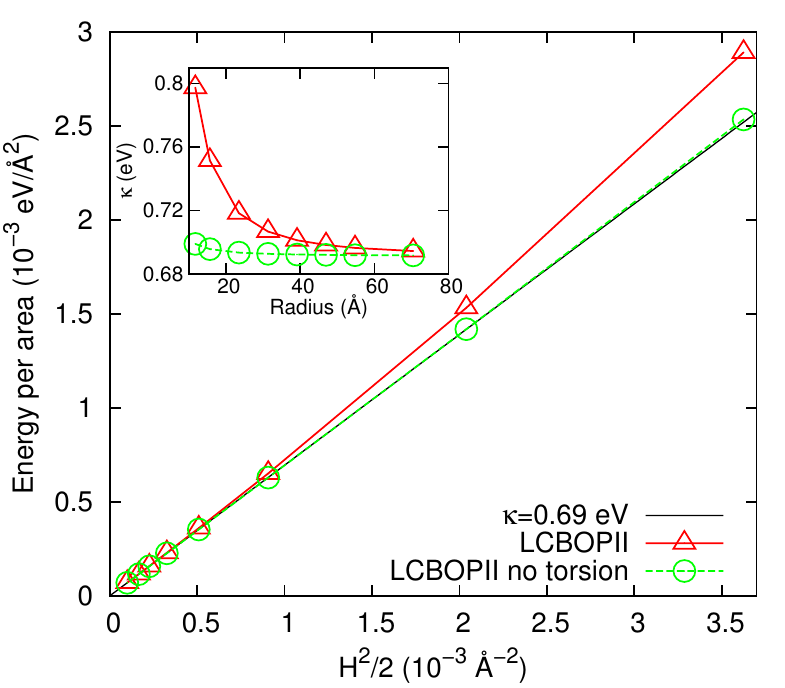}
\caption{Elastic energy per unit area, $\mathcal{E}$, versus the curvature squared, $H^2$, for nanotubes of varying radius. The slope of this curve determines the bending rigidity $\kappa$. The inset shows the bending rigidity determined for individual nanotubes.} 
\label{fig3}
\end{center}
\end{figure}

The value of 0.69 eV is low compared to other studies and also lower than the 0.82 eV reported earlier for LCBOPII at zero temperature~\cite{Fasolino2007}. This value was obtained by evaluating the elastic energy per unit area $\mathcal{E}$ of carbon nanotubes. This energy is equal to $\mathcal{E}=1/2 \kappa H^2$ where $H$ is the curvature of the nanotube. The apparent discrepancy with the result from the phonons is due to the fact that in forming a nanotube from graphene both elastic and torsional energy occur and it is only in the limit of very large nanotubes that the torsion energy can be neglected.
We calculated the elastic energy, and the corresponding bending rigidity, for several nanotubes with radii from 11 to 70 \AA\ with and without the inclusion of the torsion term. The results in Fig.~\ref{fig3} clearly demonstrate that $\mathcal{E}$ is a linear function of $H^2$, without torsion, indicating a constant $\kappa=0.69 $ eV. With the inclusion of the torsion term the resulting $\kappa$ increases with the radius of the nanotube. For large nanotubes, with small torsion angles, the value of 0.69 eV found from the phonons is recovered. The value of 0.82 eV from Ref.~\cite{Fasolino2007} was indeed determined from a nanotube with a radius of approximately 11 \AA.

\section{Nanotubes and multilayer graphene}
\label{sec:other}
\begin{figure}[h]
\begin{center}
\includegraphics{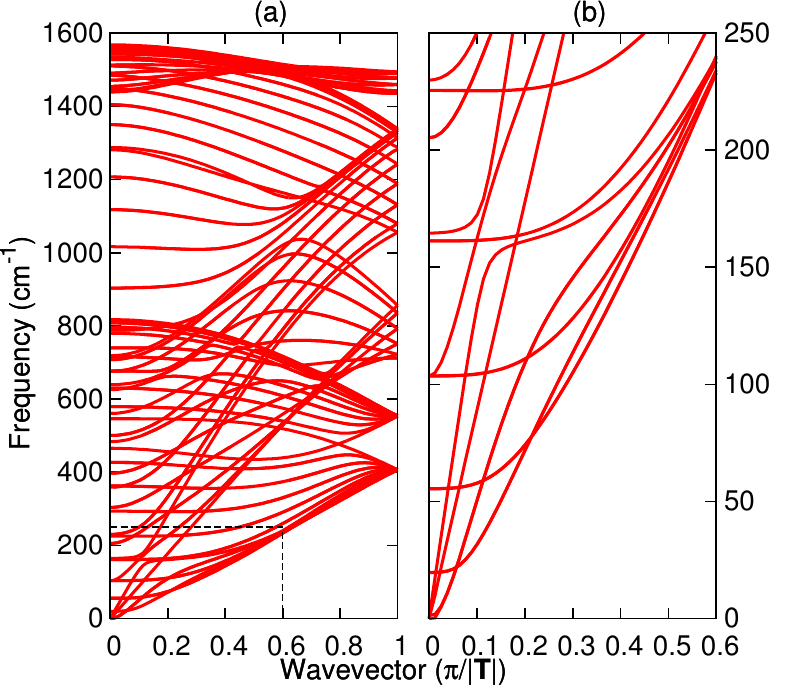}
\caption{(a) Phonon dispersion from LCBOPII for a (10,10) carbon nanotube; (b) The low frequency part of the phonon dispersion. $|\bm{T}|=2.46$~\AA\ is the lattice parameter along the nanotube axis.}
\label{fig4}
\end{center}
\end{figure}

For completeness, we show in Fig.~\ref{fig4} also the phonons of a (10,10) nanotube that can be compared to Refs.~\cite{Dresselhaus,Dubay2003}. For nanotubes the differences between models are enhanced due the complex folding of the bands. 

\begin{figure}[h]
\begin{center}
\includegraphics{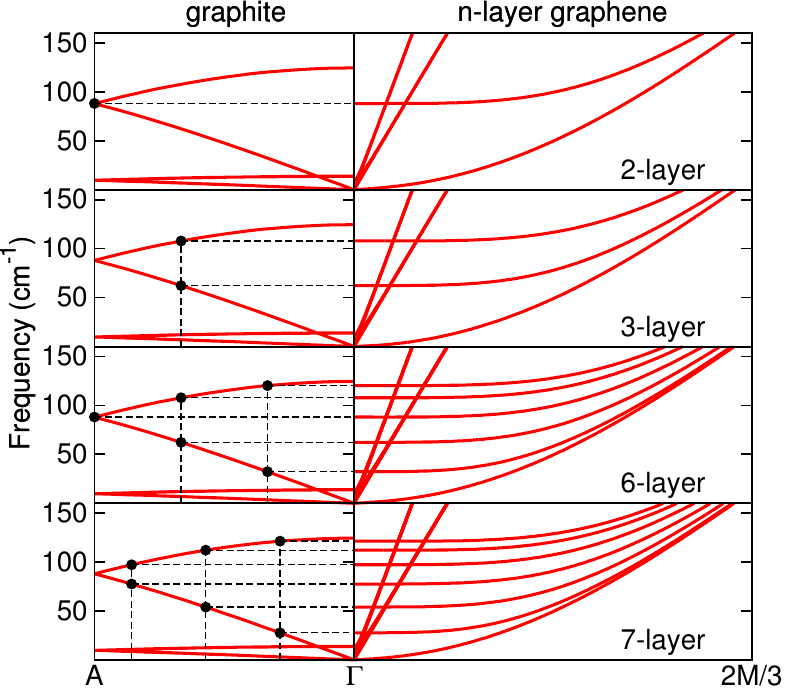}
\caption{Right panels: low frequency phonon dispersion along the $\bm{\Gamma}-\bm{M}$ line for $n$-layer graphene. Left panel: low frequency phonon dispersion along the $\bm{\Gamma}-\bm{A}$ line for graphite. The horizontal lines across the panels show that the modes of $n$-layer graphene at $\bm{\Gamma}$
coincide with those of graphene at the wavevectors given by Eq.~\eqref{Eqconf}. 
}
\label{fig5}
\end{center}
\end{figure}

Lastly, we examine the phonons of $n$-layer, $AB$ stacked, graphene, going from a single graphene layer towards graphite. In this process, the low-frequency ZA mode splits into $n$ optical sub-branches as shown in Fig.~\ref{fig5} while for all other branches the splitting is much smaller ($\sim$2~cm$^{-1}$). As shown in the left panel of Fig.~\ref{fig5}, the frequency  of these `breathing' modes at $\bm{\Gamma}$ for $n$-layer graphene are related to the longitudinal phonons of graphite along the $\bm{\Gamma}-\bm{A}$ line at wavevectors 
\begin{equation}
\bm{q}_m^n=\frac{2\pi m}{ n c}, \quad\textrm{with}\quad(m=0,\ldots,n-1),
\label{Eqconf}
\end{equation}
as if the modes were confined to an effective thickness of $nc/2$. This length is an interplanar distance larger than the actual thickness of the $n$-layer graphene. Interestingly, by extrapolating to a single layer, $n=1$, we get an effective thickness equal to an interplanar distance, as suggested in Ref.~\cite{Gupta2006}. 
Since the number and frequency of these low lying ZA modes is univocally determined by the number of graphene layers, their observation can be used for the characterization of multilayer graphene as a complement to the analysis relying on the  $2D$ band done in Refs.~\cite{Ferrari2006,Calizo2007}.

\section{Conclusions}
Empirical potentials are desirable for their simplicity and transferability to calculate the phonon frequencies of complex systems. The phonon dispersions of graphite and graphene are an important test for the accuracy of these EIPs. We have shown that LCBOPII gives good results for graphitic crystals particularly in comparison to other EIPs. We have analyzed the reasons for the remaining discrepancies, suggesting that the potential could be improved considerably by modification of the long-range interactions. The quadratic ZA bending mode plays a key role in the graphene structure at finite temperatures
and we have discussed how this fact might influence the fitting of force constants models to experimental values measured at room temperature. Lastly, we point out that multilayer graphene is characterized by several  low frequency breathing modes at $\bm{\Gamma}$, that are univocally related to the number of layers and could be used for their characterization.

\section*{Acknowledgements}
We thank Jan Los for useful discussions and Misha Katsnelson for his interest in this work. This work is part of the research program of the Stichting voor Fundamenteel Onderzoek der Materie (FOM), which is financially supported by the Nederlandse Organisatie voor Wetenschappelijk Onderzoek (NWO).


\begin{thebibliography}{48}
\expandafter\ifx\csname natexlab\endcsname\relax\def\natexlab#1{#1}\fi
\providecommand{\bibinfo}[2]{#2}
\ifx\xfnm\relax \def\xfnm[#1]{\unskip,\space#1}\fi
%Type = Article
\bibitem[{Mohr et~al.(2007)Mohr, Maultzsch, Dobard\ifmmode \check{z}\else
  \v{z}\fi{}i\ifmmode~\acute{c}\else \'{c}\fi{}, Reich, Milo\ifmmode
  \check{s}\else \v{s}\fi{}evi\ifmmode~\acute{c}\else \'{c}\fi{},
  Damnjanovi\ifmmode~\acute{c}\else \'{c}\fi{}, Bosak, Krisch, and
  Thomsen}]{Mohr2007}
\bibinfo{author}{M.~Mohr}, \bibinfo{author}{J.~Maultzsch},
  \bibinfo{author}{E.~Dobard\ifmmode \check{z}\else
  \v{z}\fi{}i\ifmmode~\acute{c}\else \'{c}\fi{}}, \bibinfo{author}{S.~Reich},
  \bibinfo{author}{I.~Milo\ifmmode \check{s}\else
  \v{s}\fi{}evi\ifmmode~\acute{c}\else \'{c}\fi{}},
  \bibinfo{author}{M.~Damnjanovi\ifmmode~\acute{c}\else \'{c}\fi{}},
  \bibinfo{author}{A.~Bosak}, \bibinfo{author}{M.~Krisch},
  \bibinfo{author}{C.~Thomsen}, \bibinfo{journal}{Phys. Rev. B}
  \bibinfo{volume}{76} (\bibinfo{year}{2007}) \bibinfo{pages}{035439}.
%Type = Article
\bibitem[{Maultzsch et~al.(2004)Maultzsch, Reich, Thomsen, Requardt, and
  Ordej\'on}]{Maultzsch2004}
\bibinfo{author}{J.~Maultzsch}, \bibinfo{author}{S.~Reich},
  \bibinfo{author}{C.~Thomsen}, \bibinfo{author}{H.~Requardt},
  \bibinfo{author}{P.~Ordej\'on}, \bibinfo{journal}{Phys. Rev. Lett.}
  \bibinfo{volume}{92} (\bibinfo{year}{2004}) \bibinfo{pages}{075501}.
%Type = Article
\bibitem[{Siebentritt et~al.(1997)Siebentritt, Pues, Rieder, and
  Shikin}]{Siebentritt1997}
\bibinfo{author}{S.~Siebentritt}, \bibinfo{author}{R.~Pues},
  \bibinfo{author}{K.-H. Rieder}, \bibinfo{author}{A.~M. Shikin},
  \bibinfo{journal}{Phys. Rev. B} \bibinfo{volume}{55} (\bibinfo{year}{1997})
  \bibinfo{pages}{7927}.
%Type = Article
\bibitem[{Nicklow et~al.(1972)Nicklow, Wakabayashi, and Smith}]{Nicklow1972}
\bibinfo{author}{R.~Nicklow}, \bibinfo{author}{N.~Wakabayashi},
  \bibinfo{author}{H.~G. Smith}, \bibinfo{journal}{Phys. Rev. B}
  \bibinfo{volume}{5} (\bibinfo{year}{1972}) \bibinfo{pages}{4951}.
%Type = Article
\bibitem[{Oshima et~al.(1988)Oshima, Aizawa, Souda, Ishizawa, and
  Sumiyoshi}]{Oshima1988}
\bibinfo{author}{C.~Oshima}, \bibinfo{author}{T.~Aizawa},
  \bibinfo{author}{R.~Souda}, \bibinfo{author}{Y.~Ishizawa},
  \bibinfo{author}{Y.~Sumiyoshi}, \bibinfo{journal}{Solid State Commun.}
  \bibinfo{volume}{65} (\bibinfo{year}{1988}) \bibinfo{pages}{1601}.
%Type = Article
\bibitem[{Tuinstra and Koenig(1970)}]{Tuinstra1970}
\bibinfo{author}{F.~Tuinstra}, \bibinfo{author}{J.~L. Koenig},
  \bibinfo{journal}{J. Chem. Phys.} \bibinfo{volume}{53} (\bibinfo{year}{1970})
  \bibinfo{pages}{1126}.
%Type = Book
\bibitem[{Saito et~al.(2007)Saito, Dresselhaus, and Dresselhaus}]{Dresselhaus}
\bibinfo{author}{R.~Saito}, \bibinfo{author}{G.~Dresselhaus},
  \bibinfo{author}{M.~Dresselhaus}, \bibinfo{title}{Physical Properties of
  Carbon Nanotubes}, \bibinfo{publisher}{Imperial College Press},
  \bibinfo{address}{London}, \bibinfo{year}{2007}.
%Type = Article
\bibitem[{Benedek and Onida(1993)}]{Benedek1993}
\bibinfo{author}{G.~Benedek}, \bibinfo{author}{G.~Onida},
  \bibinfo{journal}{Phys. Rev. B} \bibinfo{volume}{47} (\bibinfo{year}{1993})
  \bibinfo{pages}{16471}.
%Type = Article
\bibitem[{Al-Jishi and Dresselhaus(1982)}]{Al-Jishi1982}
\bibinfo{author}{R.~Al-Jishi}, \bibinfo{author}{G.~Dresselhaus},
  \bibinfo{journal}{Phys. Rev. B} \bibinfo{volume}{26} (\bibinfo{year}{1982})
  \bibinfo{pages}{4514}.
%Type = Article
\bibitem[{de~Rouffignac et~al.(1981)de~Rouffignac, Alldredge, and
  de~Wette}]{Rouffignac1981}
\bibinfo{author}{E.~de~Rouffignac}, \bibinfo{author}{G.~P. Alldredge},
  \bibinfo{author}{F.~W. de~Wette}, \bibinfo{journal}{Phys. Rev. B}
  \bibinfo{volume}{23} (\bibinfo{year}{1981}) \bibinfo{pages}{4208}.
%Type = Article
\bibitem[{Casiraghi et~al.(2009)Casiraghi, Hartschuh, Qian, Piscanec, Georgi,
  Fasoli, Novoselov, Basko, and Ferrari}]{Casiraghi2009}
\bibinfo{author}{C.~Casiraghi}, \bibinfo{author}{A.~Hartschuh},
  \bibinfo{author}{H.~Qian}, \bibinfo{author}{S.~Piscanec},
  \bibinfo{author}{C.~Georgi}, \bibinfo{author}{A.~Fasoli},
  \bibinfo{author}{K.~S. Novoselov}, \bibinfo{author}{D.~M. Basko},
  \bibinfo{author}{A.~C. Ferrari}, \bibinfo{journal}{Nano Lett.}
  \bibinfo{volume}{9} (\bibinfo{year}{2009}) \bibinfo{pages}{1433}.
%Type = Article
\bibitem[{Calizo et~al.(2007)Calizo, Balandin, Bao, Miao, and Lau}]{Calizo2007}
\bibinfo{author}{I.~Calizo}, \bibinfo{author}{A.~A. Balandin},
  \bibinfo{author}{W.~Bao}, \bibinfo{author}{F.~Miao}, \bibinfo{author}{C.~N.
  Lau}, \bibinfo{journal}{Nano Lett.} \bibinfo{volume}{7}
  (\bibinfo{year}{2007}) \bibinfo{pages}{2645}.
%Type = Article
\bibitem[{Ferrari et~al.(2006)Ferrari, Meyer, Scardaci, Casiraghi, Lazzeri,
  Mauri, Piscanec, Jiang, Novoselov, Roth, and Geim}]{Ferrari2006}
\bibinfo{author}{A.~C. Ferrari}, \bibinfo{author}{J.~C. Meyer},
  \bibinfo{author}{V.~Scardaci}, \bibinfo{author}{C.~Casiraghi},
  \bibinfo{author}{M.~Lazzeri}, \bibinfo{author}{F.~Mauri},
  \bibinfo{author}{S.~Piscanec}, \bibinfo{author}{D.~Jiang},
  \bibinfo{author}{K.~S. Novoselov}, \bibinfo{author}{S.~Roth},
  \bibinfo{author}{A.~K. Geim}, \bibinfo{journal}{Phys. Rev. Lett.}
  \bibinfo{volume}{97} (\bibinfo{year}{2006}) \bibinfo{pages}{187401}.
%Type = Article
\bibitem[{Meyer et~al.(2007)Meyer, Geim, Katsnelson, Novoselov, Booth, and
  Roth}]{Meyer2007}
\bibinfo{author}{J.~C. Meyer}, \bibinfo{author}{A.~K. Geim},
  \bibinfo{author}{M.~I. Katsnelson}, \bibinfo{author}{K.~S. Novoselov},
  \bibinfo{author}{T.~J. Booth}, \bibinfo{author}{S.~Roth},
  \bibinfo{journal}{{Nature}} \bibinfo{volume}{{446}} (\bibinfo{year}{{2007}})
  \bibinfo{pages}{{60}}.
%Type = Article
\bibitem[{Bao et~al.(2009)Bao, Miao, Chen, Zhang, Jang, Dames, and
  Lau}]{Bao2009}
\bibinfo{author}{W.~Bao}, \bibinfo{author}{F.~Miao}, \bibinfo{author}{Z.~Chen},
  \bibinfo{author}{H.~Zhang}, \bibinfo{author}{W.~Jang},
  \bibinfo{author}{C.~Dames}, \bibinfo{author}{C.~N. Lau},
  \bibinfo{journal}{Nat. Nanotechnol.} \bibinfo{volume}{4} (\bibinfo{year}{2009})
  \bibinfo{pages}{562}.
%Type = Article
\bibitem[{Zakharchenko et~al.(2009)Zakharchenko, Katsnelson, and
  Fasolino}]{Zakharchenko2009}
\bibinfo{author}{K.~V. Zakharchenko}, \bibinfo{author}{M.~I. Katsnelson},
  \bibinfo{author}{A.~Fasolino}, \bibinfo{journal}{Phys. Rev. Lett.}
  \bibinfo{volume}{102} (\bibinfo{year}{2009}) \bibinfo{pages}{046808}.
%Type = Article
\bibitem[{Koskinen et~al.(2009)Koskinen, Malola, and H\"akkinen}]{Koskinen2009}
\bibinfo{author}{P.~Koskinen}, \bibinfo{author}{S.~Malola},
  \bibinfo{author}{H.~H\"akkinen}, \bibinfo{journal}{Phys. Rev. B}
  \bibinfo{volume}{80} (\bibinfo{year}{2009}) \bibinfo{pages}{073401}.
%Type = Article
\bibitem[{Carlsson and Scheffler(2006)}]{Carlsson2006}
\bibinfo{author}{J.~M. Carlsson}, \bibinfo{author}{M.~Scheffler},
  \bibinfo{journal}{Phys. Rev. Lett.} \bibinfo{volume}{96}
  (\bibinfo{year}{2006}) \bibinfo{pages}{046806}.
%Type = Article
\bibitem[{Lahiri et~al.(2010)Lahiri, Lin, Bozkurt, Oleynik, and
  Batzill}]{Lahiri2010}
\bibinfo{author}{J.~Lahiri}, \bibinfo{author}{Y.~Lin},
  \bibinfo{author}{P.~Bozkurt}, \bibinfo{author}{I.~I. Oleynik},
  \bibinfo{author}{M.~Batzill}, \bibinfo{journal}{{Nat. Nanotechnol.}}
  \bibinfo{volume}{{5}} (\bibinfo{year}{{2010}}) \bibinfo{pages}{{326}}.
%Type = Article
\bibitem[{Coraux et~al.(2008)Coraux, N'Diaye, Busse, and Michely}]{Coraux2008}
\bibinfo{author}{J.~Coraux}, \bibinfo{author}{A.~T. N'Diaye},
  \bibinfo{author}{C.~Busse}, \bibinfo{author}{T.~Michely},
  \bibinfo{journal}{{Nano Lett.}} \bibinfo{volume}{{8}}
  (\bibinfo{year}{{2008}}) \bibinfo{pages}{{565}}.
%Type = Article
\bibitem[{Los and Fasolino(2003)}]{Los2003}
\bibinfo{author}{J.~H. Los}, \bibinfo{author}{A.~Fasolino},
  \bibinfo{journal}{Phys. Rev. B} \bibinfo{volume}{68} (\bibinfo{year}{2003})
  \bibinfo{pages}{024107}.
%Type = Article
\bibitem[{Brenner(1990)}]{Brenner1990}
\bibinfo{author}{D.~W. Brenner}, \bibinfo{journal}{Phys. Rev. B}
  \bibinfo{volume}{42} (\bibinfo{year}{1990}) \bibinfo{pages}{9458}.
%Type = Article
\bibitem[{Tersoff(1986)}]{Tersoff1986}
\bibinfo{author}{J.~Tersoff}, \bibinfo{journal}{Phys. Rev. Lett.}
  \bibinfo{volume}{56} (\bibinfo{year}{1986}) \bibinfo{pages}{632}.
%Type = Article
\bibitem[{Fasolino et~al.(2007)Fasolino, Los, and Katsnelson}]{Fasolino2007}
\bibinfo{author}{A.~Fasolino}, \bibinfo{author}{J.~H. Los},
  \bibinfo{author}{M.~I. Katsnelson}, \bibinfo{journal}{Nat. Mater.}
  \bibinfo{volume}{6} (\bibinfo{year}{2007}) \bibinfo{pages}{858}.
%Type = Article
\bibitem[{Glosli and Ree(1999)}]{Glosli1999}
\bibinfo{author}{J.~N. Glosli}, \bibinfo{author}{F.~H. Ree},
  \bibinfo{journal}{Phys. Rev. Lett.} \bibinfo{volume}{82}
  (\bibinfo{year}{1999}) \bibinfo{pages}{4659}.
%Type = Article
\bibitem[{Ghiringhelli et~al.(2005)Ghiringhelli, Los, Meijer, Fasolino, and
  Frenkel}]{Ghiringhelli2005}
\bibinfo{author}{L.~M. Ghiringhelli}, \bibinfo{author}{J.~H. Los},
  \bibinfo{author}{E.~J. Meijer}, \bibinfo{author}{A.~Fasolino},
  \bibinfo{author}{D.~Frenkel}, \bibinfo{journal}{Phys. Rev. Lett.}
  \bibinfo{volume}{94} (\bibinfo{year}{2005}) \bibinfo{pages}{145701}.
%Type = Article
\bibitem[{Los et~al.(2005)Los, Ghiringhelli, Meijer, and Fasolino}]{Los2005}
\bibinfo{author}{J.~H. Los}, \bibinfo{author}{L.~M. Ghiringhelli},
  \bibinfo{author}{E.~J. Meijer}, \bibinfo{author}{A.~Fasolino},
  \bibinfo{journal}{Phys. Rev. B} \bibinfo{volume}{72} (\bibinfo{year}{2005})
  \bibinfo{pages}{214102}.
%Type = Article
\bibitem[{Tersoff(1988)}]{Tersoff1988}
\bibinfo{author}{J.~Tersoff}, \bibinfo{journal}{Phys. Rev. Lett.}
  \bibinfo{volume}{61} (\bibinfo{year}{1988}) \bibinfo{pages}{2879}.
%Type = Article
\bibitem[{Yanagisawa et~al.(2005)Yanagisawa, Tanaka, Ishida, Matsue, Rokuta,
  Otani, and Oshima}]{Yanagisawa2005}
\bibinfo{author}{H.~Yanagisawa}, \bibinfo{author}{T.~Tanaka},
  \bibinfo{author}{Y.~Ishida}, \bibinfo{author}{M.~Matsue},
  \bibinfo{author}{E.~Rokuta}, \bibinfo{author}{S.~Otani},
  \bibinfo{author}{C.~Oshima}, \bibinfo{journal}{Surf. Interface Anal.}
  \bibinfo{volume}{{37}} (\bibinfo{year}{{2005}}) \bibinfo{pages}{{133}}.
%Type = Article
\bibitem[{Mounet and Marzari(2005)}]{Mounet2005}
\bibinfo{author}{N.~Mounet}, \bibinfo{author}{N.~Marzari},
  \bibinfo{journal}{Phys. Rev. B} \bibinfo{volume}{71} (\bibinfo{year}{2005})
  \bibinfo{pages}{205214}.
%Type = Article
\bibitem[{Wirtz and Rubio(2004)}]{Wirtz2004}
\bibinfo{author}{L.~Wirtz}, \bibinfo{author}{A.~Rubio}, \bibinfo{journal}{Solid
  State Commun.} \bibinfo{volume}{131} (\bibinfo{year}{2004})
  \bibinfo{pages}{141}.
%Type = Article
\bibitem[{Dubay and Kresse(2003)}]{Dubay2003}
\bibinfo{author}{O.~Dubay}, \bibinfo{author}{G.~Kresse},
  \bibinfo{journal}{Phys. Rev. B} \bibinfo{volume}{67} (\bibinfo{year}{2003})
  \bibinfo{pages}{035401}.
%Type = Article
\bibitem[{Lindsay and Broido(2010)}]{Lindsay2010}
\bibinfo{author}{L.~Lindsay}, \bibinfo{author}{D.~A. Broido},
  \bibinfo{journal}{Phys. Rev. B} \bibinfo{volume}{81} (\bibinfo{year}{2010})
  \bibinfo{pages}{205441}.
%Type = Article
\bibitem[{Tewary and Yang(2009)}]{Tewary2009}
\bibinfo{author}{V.~K. Tewary}, \bibinfo{author}{B.~Yang},
  \bibinfo{journal}{Phys. Rev. B} \bibinfo{volume}{79} (\bibinfo{year}{2009})
  \bibinfo{pages}{075442}.
%Type = Book
\bibitem[{Born and Huang(1956)}]{Born}
\bibinfo{author}{M.~Born}, \bibinfo{author}{K.~Huang},
  \bibinfo{title}{Dynamical Theory of Crystal Lattices},
  \bibinfo{publisher}{Oxford University Press}, \bibinfo{year}{1956}.
%Type = Article
\bibitem[{Lifshitz(1952)}]{Lifshitz1952}
\bibinfo{author}{I.~M. Lifshitz}, \bibinfo{journal}{Zh. Eksp. Teor. Fiz.}
  \bibinfo{volume}{22} (\bibinfo{year}{1952}) \bibinfo{pages}{475}.
%Type = Article
\bibitem[{Zabel(2001)}]{Zabel2001}
\bibinfo{author}{H.~Zabel}, \bibinfo{journal}{J. Phys. Condens. Matter}
  \bibinfo{volume}{13} (\bibinfo{year}{2001}) \bibinfo{pages}{7679}.
%Type = Article
\bibitem[{Nika et~al.(2009)Nika, Pokatilov, Askerov, and Balandin}]{Nika2009}
\bibinfo{author}{D.~L. Nika}, \bibinfo{author}{E.~P. Pokatilov},
  \bibinfo{author}{A.~S. Askerov}, \bibinfo{author}{A.~A. Balandin},
  \bibinfo{journal}{Phys. Rev. B} \bibinfo{volume}{79} (\bibinfo{year}{2009})
  \bibinfo{pages}{155413}.
%Type = Article
\bibitem[{Bosak et~al.(2007)Bosak, Krisch, Mohr, Maultzsch, and
  Thomsen}]{Bosak2007}
\bibinfo{author}{A.~Bosak}, \bibinfo{author}{M.~Krisch},
  \bibinfo{author}{M.~Mohr}, \bibinfo{author}{J.~Maultzsch},
  \bibinfo{author}{C.~Thomsen}, \bibinfo{journal}{Phys. Rev. B}
  \bibinfo{volume}{75} (\bibinfo{year}{2007}) \bibinfo{pages}{153408}.
%Type = Article
\bibitem[{Piscanec et~al.(2004)Piscanec, Lazzeri, Mauri, Ferrari, and
  Robertson}]{Piscanec2004}
\bibinfo{author}{S.~Piscanec}, \bibinfo{author}{M.~Lazzeri},
  \bibinfo{author}{F.~Mauri}, \bibinfo{author}{A.~C. Ferrari},
  \bibinfo{author}{J.~Robertson}, \bibinfo{journal}{Phys. Rev. Lett.}
  \bibinfo{volume}{93} (\bibinfo{year}{2004}) \bibinfo{pages}{185503}.
%Type = Book
\bibitem[{Nelson et~al.(2004)Nelson, Piran, and Weinberg}]{Nelson}
\bibinfo{editor}{D.~Nelson}, \bibinfo{editor}{T.~Piran},
  \bibinfo{editor}{S.~Weinberg} (Eds.), \bibinfo{title}{Statistical Mechanics
  of Membranes and Surfaces}, \bibinfo{publisher}{World Scientific},
  \bibinfo{address}{Singapore}, \bibinfo{year}{2004}.
%Type = Article
\bibitem[{Zakharchenko et~al.(2010)Zakharchenko, Los, Katsnelson, and
  Fasolino}]{Zakharchenko2010}
\bibinfo{author}{K.~V. Zakharchenko}, \bibinfo{author}{J.~H. Los},
  \bibinfo{author}{M.~I. Katsnelson}, \bibinfo{author}{A.~Fasolino},
  \bibinfo{journal}{Phys. Rev. B} \bibinfo{volume}{81} (\bibinfo{year}{2010})
  \bibinfo{pages}{235439}.
%Type = Article
\bibitem[{Tersoff(1992)}]{Tersoff1992}
\bibinfo{author}{J.~Tersoff}, \bibinfo{journal}{Phys. Rev. B}
  \bibinfo{volume}{46} (\bibinfo{year}{1992}) \bibinfo{pages}{15546}.
%Type = Article
\bibitem[{Perebeinos and Tersoff(2009)}]{Perebeinos2009}
\bibinfo{author}{V.~Perebeinos}, \bibinfo{author}{J.~Tersoff},
  \bibinfo{journal}{Phys. Rev. B} \bibinfo{volume}{79} (\bibinfo{year}{2009})
  \bibinfo{pages}{241409}.
%Type = Article
\bibitem[{Tu and Ou-Yang(2002)}]{Zhanchun2002}
\bibinfo{author}{Z.-c. Tu}, \bibinfo{author}{Z.-c. Ou-Yang},
  \bibinfo{journal}{Phys. Rev. B} \bibinfo{volume}{65} (\bibinfo{year}{2002})
  \bibinfo{pages}{233407}.
%Type = Article
\bibitem[{Kolmogorov and Crespi(2005)}]{Kolmogorov2005}
\bibinfo{author}{A.~N. Kolmogorov}, \bibinfo{author}{V.~H. Crespi},
  \bibinfo{journal}{Phys. Rev. B} \bibinfo{volume}{71} (\bibinfo{year}{2005})
  \bibinfo{pages}{235415}.
%Type = Article
\bibitem[{Savini et~al.(2010)Savini, Dappe, {\"{O}}berg, Charlier, Katsnelson,
  and Fasolino}]{Savini2010}
\bibinfo{author}{G.~Savini}, \bibinfo{author}{Y.~Dappe},
  \bibinfo{author}{S.~{\"{O}}berg}, \bibinfo{author}{J.-C. Charlier},
  \bibinfo{author}{M.~Katsnelson}, \bibinfo{author}{A.~Fasolino},
  \bibinfo{journal}{Carbon} \bibinfo{volume}{In Press, Accepted Manuscript}
  (\bibinfo{year}{2010}) \bibinfo{pages}{--}.
%Type = Article
\bibitem[{Gupta et~al.(2006)Gupta, Chen, Joshi, Tadigadapa, and
  Eklund}]{Gupta2006}
\bibinfo{author}{A.~Gupta}, \bibinfo{author}{G.~Chen},
  \bibinfo{author}{P.~Joshi}, \bibinfo{author}{S.~Tadigadapa},
  \bibinfo{author}{Eklund}, \bibinfo{journal}{Nano Lett.} \bibinfo{volume}{6}
  (\bibinfo{year}{2006}) \bibinfo{pages}{2667}.

\end{thebibliography}
\end{document}